\documentclass[conference]{IEEEtran}
\IEEEoverridecommandlockouts

\usepackage{cite}
\usepackage{amsmath,amssymb,amsfonts}
\usepackage{algorithmic}
\usepackage{graphicx}
\usepackage{textcomp}
\usepackage{booktabs}
\usepackage{float}
\usepackage{url}  
\usepackage{cuted}
\usepackage{caption}
\usepackage{orcidlink}

\usepackage{xcolor}
\newcommand{\code}[1]{\colorbox{gray!15}{\texttt{#1}}}

\def\BibTeX{{\rm B\kern-.05em{\sc i\kern-.025em b}\kern-.08em
    T\kern-.1667em\lower.7ex\hbox{E}\kern-.125emX}}
\begin{document}

\title{TUMSphere: Turning a University Curriculum into Playable VR Challenges}

\author{
\IEEEauthorblockN{
\begin{tabular}[t]{ccc}
 Santiago Berrezueta-Guzman\,\orcidlink{0000-0001-5559-2056} &
 Nadia Damianova\,\orcidlink{0009-0002-6945-0870}&
 Andrei Koshelev\,\orcidlink{0009-0000-3786-4994}\\
\textit{\small Technical University of Munich} &
\textit{\small Technical University of Munich} &
\textit{\small Technical University of Munich} \\
\small Heilbronn, Germany &
\small Heilbronn, Germany &
\small Heilbronn, Germany \\
[2ex]
\multicolumn{3}{c}{
\begin{tabular}[t]{cc}
Ivan Parmacli\,\orcidlink{0009-0004-1575-4478}&
Stefan Wagner\,\orcidlink{0000-0002-5256-8429}\\
\textit{\small Technical University of Munich} &
\textit{\small Technical University of Munich} \\
\small Heilbronn, Germany &
\small Heilbronn, Germany \\
\end{tabular}
}
\end{tabular}
}
}

\maketitle

\begin{abstract}

Traditional university orientation formats struggle to convey the intellectual substance of STEM curricula, particularly in disciplines where core competencies, such as algorithmic thinking and formal reasoning, are inherently abstract. This paper presents \textbf{TUMSphere}, a serious virtual reality (VR) application built as an interactive digital twin of the TUM Bildungscampus Heilbronn, in which six curriculum-mapped mini-games translate foundational Information Engineering topics into hands-on VR challenges. The mini-games—covering introductory programming, hardware debugging, code completion, graph traversal, shortest-path optimization, and relational database querying—follow a graduated difficulty progression that mirrors the real semesters' structure of the degree. 

We describe the pedagogical rationale, the VR interaction mechanics, and nine cross-cutting design considerations that guided development. A within-subjects pilot study ($N = 18$) using pre-/post-knowledge tests, the System Usability Scale, a User Engagement Scale adaptation, and the Simulator Sickness Questionnaire yielded a statistically significant knowledge gain ($p < 0.001$, $r = 0.86$), good usability (SUS $M = 76.4$), high engagement ($M = 4.21/5$), and negligible simulator sickness (SSQ $M = 7.1$). Task performance logs confirmed the intended difficulty gradient across mini-games. These results suggest that embedding authentic academic challenges in an explorable VR campus is a viable and extensible approach to gamified STEM outreach.

\end{abstract}


\section{Introduction}\label{I}

University orientation programs and open-campus days typically rely on guided walks, printed brochures, and slide-based presentations to convey what a degree program entails. While these formats communicate logistical information effectively, they offer prospective students little opportunity to experience the intellectual substance of a curriculum first-hand \cite{korstange2020online}. The problem is particularly acute in STEM disciplines such as Information Engineering, where core competencies such as algorithmic thinking, formal reasoning about data structures, and systematic debugging are inherently abstract and difficult to convey through passive media \cite{winberg2019learning, hasanah2020key}. 

At the same time, a growing body of research shows that immersive technologies can bridge this gap: virtual reality environments enable learners to interact physically with conceptual representations, turning observation into action and significantly improving both motivation and retention \cite{geng2021application, ogbuanya2018investigating}. Therefore, \textit{TUMSphere}\footnote{Project website: \url{https://tumsphere.se.cit.tum.de/}.} was conceived to exploit this potential by embedding authentic academic challenges directly into an explorable VR replica of the Technical University of Munich (TUM) campus in Heilbronn, giving visitors a hands-on preview of the skills they will develop throughout the degree \cite{damianova2026anatomy}.

A central design goal of TUMSphere was to move beyond passive campus exploration and instead offer players a structured, playful encounter with the academic material taught in the Information Engineering program at TUM Heilbronn. Rather than presenting course descriptions as static text, the development team chose to encode key learning objectives into a sequence of interactive mini-games. This approach draws on evidence that experiential interaction with abstract subject matter can improve both engagement and conceptual retention, particularly in STEM contexts.

The mini-games form a coherent progression that shadows the real semester structure of the degree. Early challenges address foundational topics such as syntax and basic logic, while later ones introduce more advanced ideas, such as relational databases and combinatorial search. 

The remainder of this paper is organized as follows: Section~\ref{sec:rw} reviews related work and compares with our development, Section~\ref{sec:flow} describes the overall game flow, Section~\ref{sec:minigames} details each mini-game and its pedagogical intent, Section~\ref{sec:design} discusses cross-cutting design considerations, Section~\ref{sec:eval} presents the pilot evaluation methodology and section~\ref{sec:results} its results. Section~\ref{sec:conclusion} concludes the paper with observations, and section \ref{sec:future} provides future directions.

\section{Related Work}\label{sec:rw}

Recent literature emphasizes that the metaverse offers transformative potential for educational technology by providing immersive learning experiences that transcend physical classroom limitations \cite{lin2022metaverse}. 

Putri et al.\ identified the metaverse as a significant innovation for curriculum development, enabling a shift from static, traditional curricula to more flexible, interactive designs that foster creativity, critical thinking, and collaborative skills. Their study highlighted that integrating such technology enables experience-based learning, where students ``feel'' rather than just read, thereby improving the absorption of complex materials~\cite{putri2024use}.

Programming education continues to face significant challenges, including high dropout rates and a persistent gap between students' natural reasoning and the computer-oriented logic required for software development. Konecki et al.\ demonstrate that hands-on VR experiences significantly improve students' perceived motivation and their belief that they would understand complex programming concepts, such as memory allocation and linked lists, better through interactive visualization~\cite{konecki2023using}.

Pirker et al.\ showed that VR implementations of algorithm visualizations induce higher levels of presence, absorption, flow, and psychological immersion compared to traditional desktop-based web applications. Their study highlights that learners in a VR setting report stronger positive emotions and perceive the experience as more engaging and motivating. Spatial metaphors, such as physically grabbing "algorithm cubes" or interacting with spheres representing data elements, make abstract phenomena more concrete for students. Despite these benefits, web-based versions remain valuable for large-scale classroom settings, suggesting that VR is most effective as a supplementary tool for personalized or small-group experiential learning~\cite{pirker2021potential}.

Table~\ref{tab:comparison} positions \textit{TUMSphere} relative to these works. While existing research validates VR's superiority in fostering engagement and absorption, many implementations remain limited to isolated scenarios or general frameworks. \textit{TUMSphere} differentiates itself by translating abstract STEM topics, such as relational filtering and graph traversal, into subject-specific VR mechanics arranged in a structured progression that mirrors the real semester sequence.

\begin{strip}
\centering
\captionof{table}{Comparison of Related Works and the TUMSphere Framework}
\label{tab:comparison}
\begin{tabular}{@{}p{1.8cm}p{3.6cm}p{3.4cm}p{3.7cm}p{4.1cm}@{}}
\toprule
\textbf{Aspect} & \textbf{Putri et al.\ (2024)} & \textbf{Konecki et al.\ (2023)} & \textbf{Pirker et al.\ (2021)} & \textbf{TUMSphere (Ours)} \\ \midrule
\textbf{Primary Focus} & Metaverse as curriculum innovation. & Programming education challenges and VR. & Visualization of sorting algorithms. & Holistic IE curriculum mapping. \\\midrule
\textbf{Methodology} & Qualitative literature study and analysis. & Questionnaire and custom VR lesson. & A/B study comparing WebGL vs.\ VR. & Serialized mini-games mirroring semester progression. \\\midrule
\textbf{Core Objective} & Driving curriculum progress and creativity. & Fostering problem-solving and algorithms. & Enhancing presence, absorption, and flow. & Translating STEM theory into tactile VR mechanics. \\\midrule
\textbf{Key Barrier} & Limited technological infrastructure. & High cost and content complexity. & Cognitive overload or cybersickness. & Calibrating task complexity for spatial reasoning. \\\midrule
\textbf{Contribution} & Theoretical framework for metaverse integration. & Improving student motivation toward VR. & Proving VR's engagement superiority over the web. & Serialized academic journey. \\ 
\bottomrule
\end{tabular}
\end{strip}

\section{Overall Game Flow and Progression}\label{sec:flow}

\textit{TUMSphere} is developed using Unreal Engine~5 (UE5) for its Blueprint visual scripting system, which enables rapid iteration without deep C++ expertise; its first-class XR support via the OpenXR API and MetaXR plugin, which provides ready-made components for stereoscopic rendering, motion tracking, and object interaction; and its high-fidelity rendering pipeline, which allowed the team to produce a visually convincing digital twin of the campus that reinforces the sense of presence central to the pedagogical experience \cite{berrezueta2026choosing}. 

The development followed an academically adapted Scrum process in
which an eight-person student team used one-week sprints, a
Notion-managed product backlog, and hybrid on-site/remote
coordination via Discord and GitHub over the course of one
year~\cite{damianova2026anatomy}.

Its core gameplay is structured around curriculum-driven mini-games that translate abstract academic concepts into interactive, tactile VR mechanics. To enhance social immersion, the environment incorporates intelligent Non-Player Characters (NPCs) powered by Large Language Models, enabling real-time conversational interaction with virtual campus guides such as \textit{Akira} ~\cite{berrezueta8next}.

\begin{figure}[h!]
    \centering
    \includegraphics[width=1\linewidth]{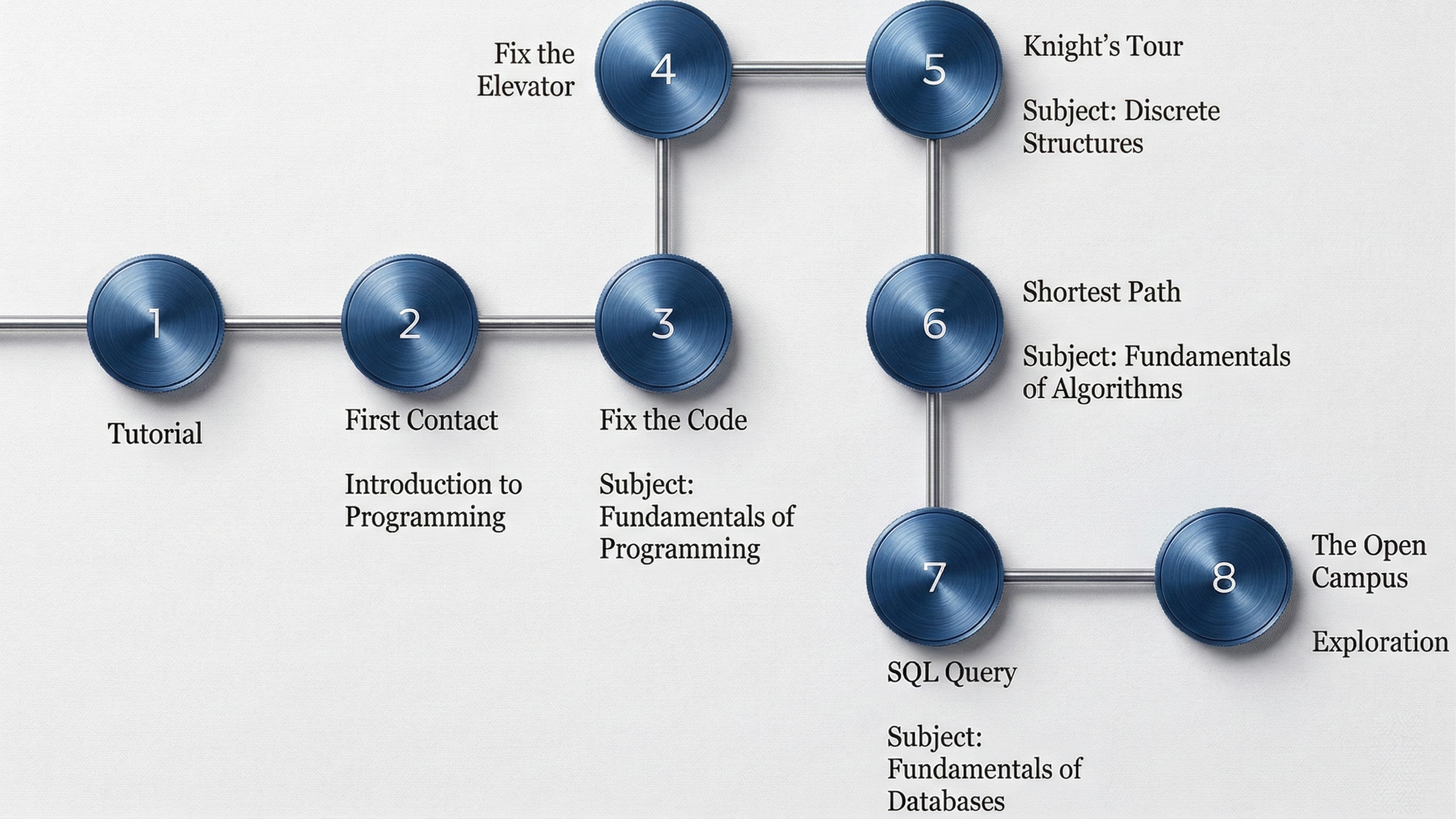}
    \caption{Eight-stage roadmap of the Information Engineering interactive study course, featuring dark blue progress nodes on a clean white background.}
    \label{map}
\end{figure}

Figure \ref{map} illustrates that the game begins in a confined starting room where a floating robot companion named \textbf{TUMi} introduces the fundamental VR controls through a guided tutorial. 
After completing the tutorial, the player advances through a series of mini-games arranged in order of rising difficulty, mirroring the distribution of courses across semesters: early-semester subjects such as computer architecture and introductory programming come first, followed by mid-program topics such as databases and discrete mathematics. Completing all challenges unlocks free exploration of the entire campus. An animated question mark icon placed near each playable zone signals that a mini-game is available at that location.

\section{Mini-Games}\label{sec:minigames}

\subsection{The Tutorial}\label{sec:tutorial}

\begin{figure}[h!]
    \centering
    \includegraphics[width=\linewidth]{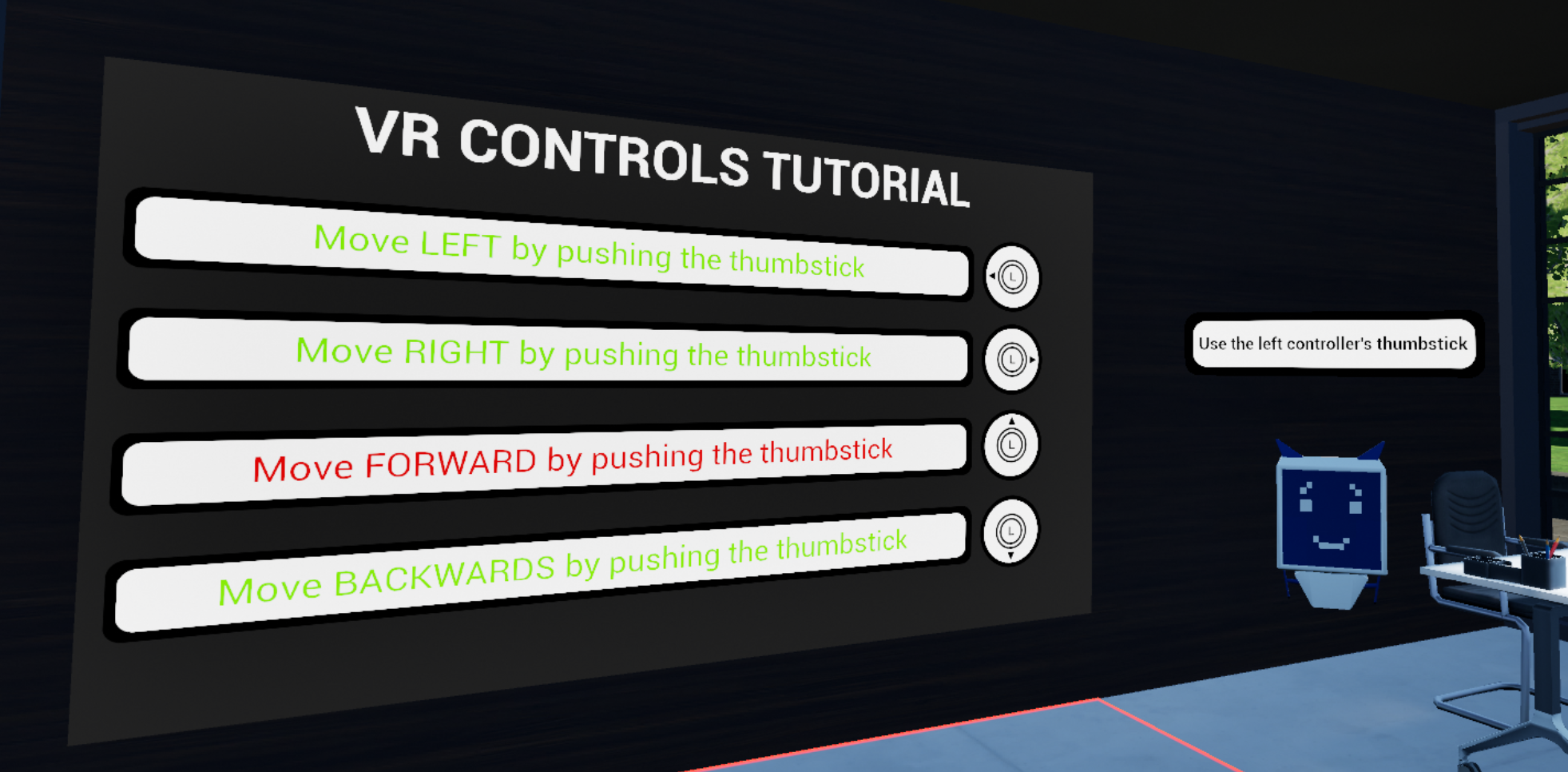}
    \caption{The tutorial room where TUMi guides the player through basic VR controls, before starting the building exploration and the rest of the mini games.}
    \label{fig:tutorial}
\end{figure}

This mini-game serves as the onboarding stage and establishes the player's VR literacy. Accordingly, before any academic content is introduced, \textit{TUMi} delivers step-by-step instructions in the Student Service Office for each control action (walking, turning, teleporting, and grabbing objects). Visual indicators start in red and turn green once the player successfully performs the corresponding input. Only after all indicators turn green (as illustrated in Figure \ref{fig:tutorial}) does the door open, granting access to the building's lobby. 

\textit{Learning outcome:} This gating mechanism ensures that players possess the motor vocabulary needed for all subsequent challenges.

\subsection{First Contact -- Hello, World! }

\begin{figure}[h!]
    \centering
    \includegraphics[width=\linewidth]{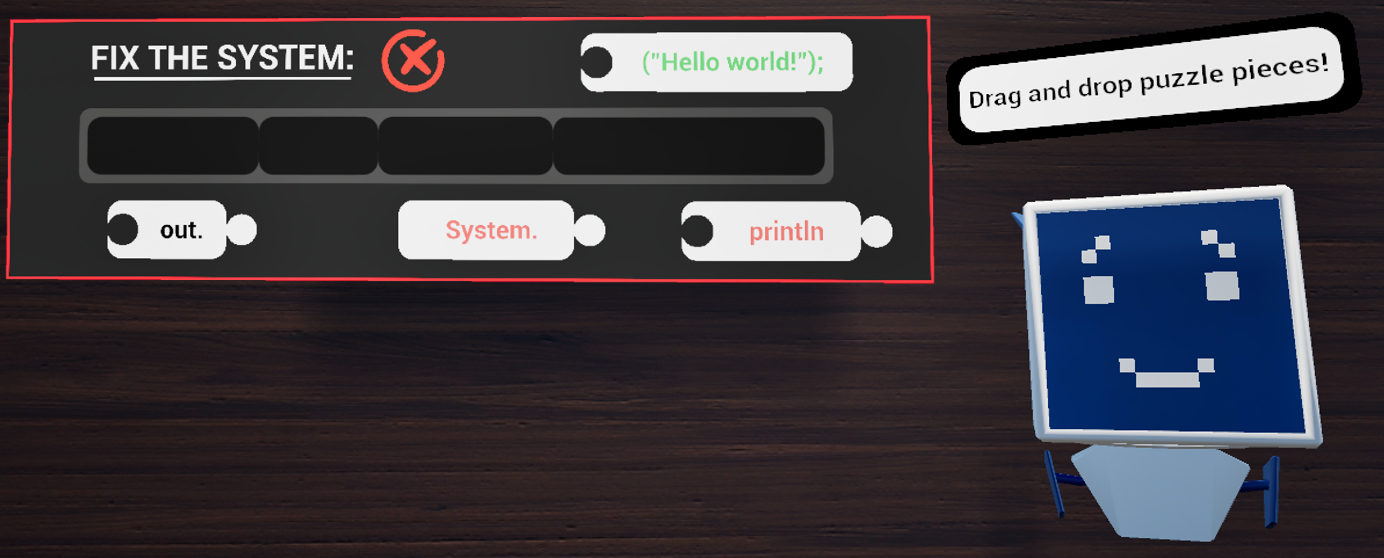}
    \caption{The First Contact mini-game: print your first \code{Hello, World!} statement from word blocks.}
    \label{fig:firstcontact}
\end{figure}

This mini-game is associated with the Introduction to Programming course. In this opening mini-game, \textit{TUMi} asks the player to physically arrange word pieces into the correct sequence to form the classic \code{System.out.println("Hello, World!");} statement as illustrated in Figure \ref{fig:firstcontact}. Using the VR controllers, the player grabs individual pieces (keywords, punctuation, and string literals) and snaps them into a designated order on a virtual display.

When the player does it correctly, there's an immediate confetti explosion around, and \textit{TUMi} indicates that the player should proceed to the next task. If the player fails, the feedback provides an intuitive hint that prompts them to reconsider the element block order until they get it right. 

\textit{Learning outcome:} Players gain an intuitive sense for how programming statements are structured, including the importance of syntax order and delimiters. Turning code composition into a tangible, spatial task reduces the intimidation factor that text-based environments can pose for beginners.

\subsection{Fix The Elevator -- Rewire!}

\begin{figure}[h!]
    \centering
    \includegraphics[width=\linewidth]{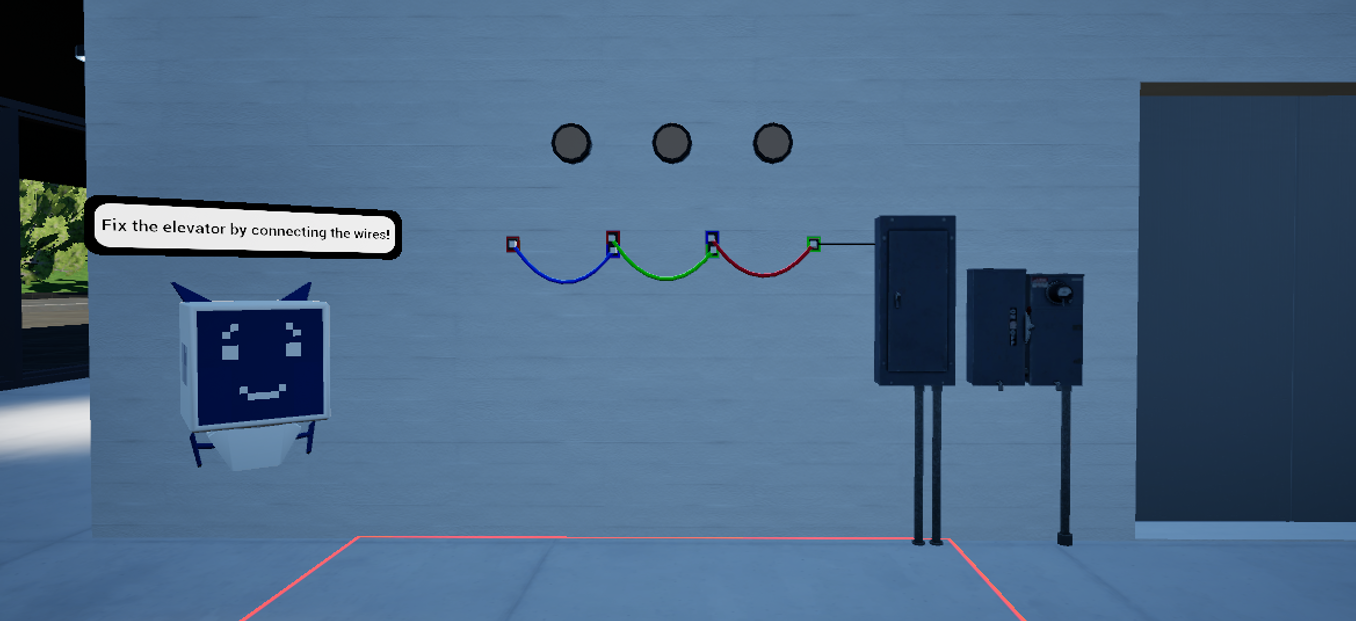}
    \caption{The Fix The Elevator mini-game: reconnecting mismatched cables to colored sockets.}
    \label{fig:elevator}
\end{figure}

The building's elevator is out of service because three colored cables (red, green, and blue) are connected to the wrong terminals. As illustrated in Figure \ref{fig:elevator}, the player must grab each cable and drag it to the matching socket. Immediate visual feedback — such as a glow effect — confirms a correct connection. Once all three wires are properly seated, the elevator becomes functional for the remainder of the game, granting vertical access to every floor.

\textit{Learning outcome:} Although simplified, the task mirrors the logic of hardware debugging: identifying mismatched connections and restoring correct signal paths. It also serves a practical game-design purpose by unlocking a navigation shortcut as a reward. This minigame also serves as a first approach for implementing hardware repair and development tasks in other subjects, such as computer networks and telecommunications. 

\subsection{Fix The Code -- Bug Hunt!}

\begin{figure}[h!]
    \centering
    \includegraphics[width=\linewidth]{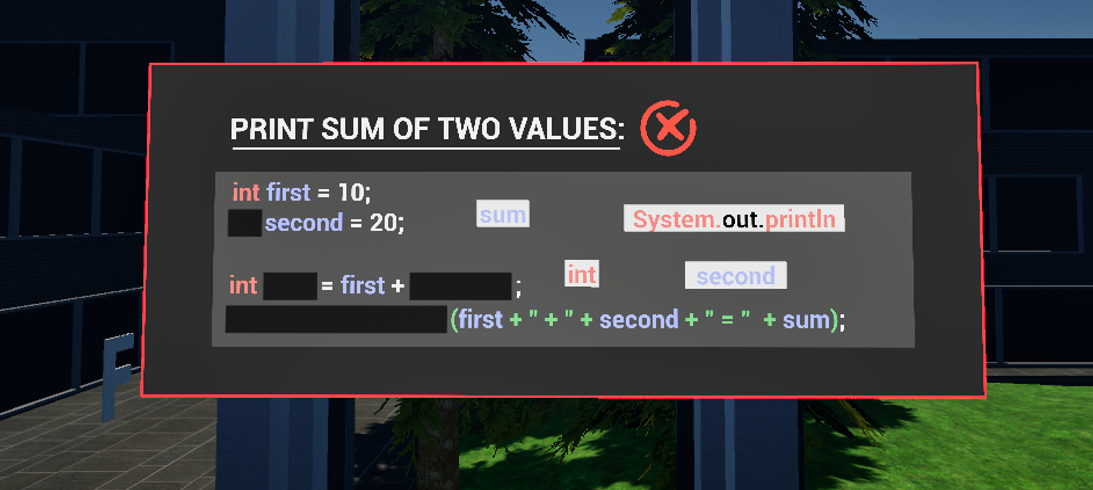}
    \caption{The Fix The Code mini-game: filling in missing tokens in a program.}
    \label{fig:fixcode}
\end{figure}

This challenge presents a partially completed program on a virtual screen. In this case, the code declares two integer variables and computes and displays their sum. Still, several parts are missing — variable names, arithmetic operators, and parts of the print statement. As illustrated in Figure \ref{fig:fixcode}, the player must select the correct identifiers and symbols from a set of options and place them into the blank regions. A correctly completed program would, \code{int sum = first + second;} and output a result such as \code{10 + 20 = 30}.

\textit{Learning outcome:} The activity reinforces variable declaration, assignment, arithmetic operations, and console output. Requiring the player to reason about which token belongs where develops the same debugging mindset that students need when reading and correcting real source code.

\subsection{Knight's Tour}

The Knight's Tour is a classical combinatorial puzzle that asks whether a chess knight can visit every square on a board exactly once using only its L-shaped move---a problem closely related to finding a Hamiltonian path in graph theory \cite{bai2006generalized}.

\begin{figure}[h!]
    \centering
    \includegraphics[width=\linewidth]{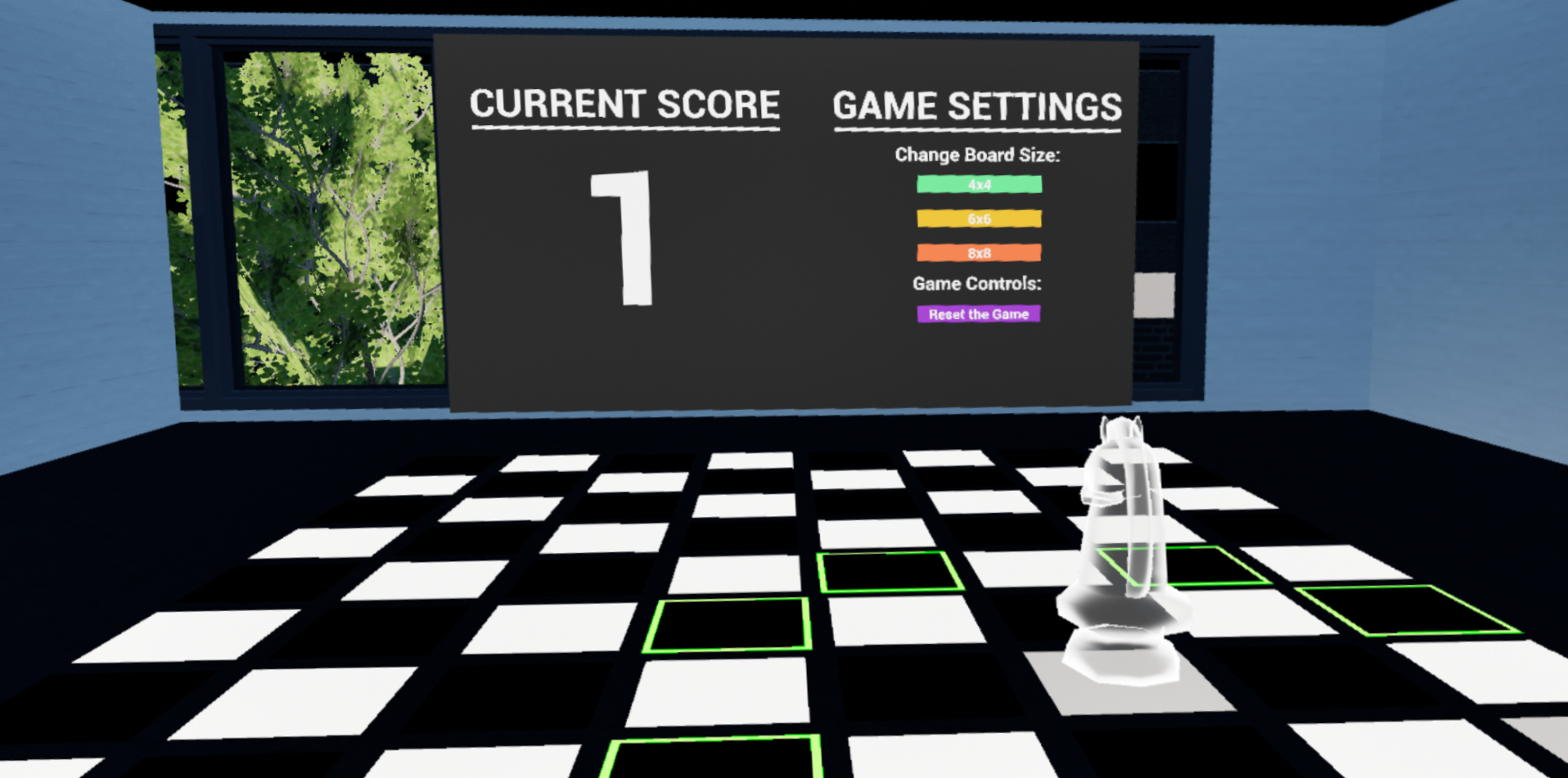}
    \caption{The Knight's Tour mini-game on a virtual chessboard.}
    \label{fig:knight}
\end{figure}

The player is presented with a configurable chessboard (8$\times$8). As illustrated in Figure \ref{fig:knight}, the virtual knight starts in one corner, and all currently legal moves are highlighted in green. The player selects the next square by pointing with the VR controller, at which point the knight relocates, the visited tile is marked, and new valid destinations are recalculated. 

A running score tracks successful moves, and the board can be reset at any time to attempt a different strategy.

\textit{Learning outcome:} Players engage with graph traversal and path-finding in a concrete, visual way. Experimenting with different move sequences and observing dead ends builds intuition for vertex connectivity and backtracking---ideas that recur throughout courses on algorithms and discrete mathematics. This mini-game is a good introduction to the upcoming high-semester lectures, such as Discrete Structures and Fundamentals of Algorithms.

\subsection{Shortest Path -- Dijkstra's Dilemma}

\begin{figure}[h!]
    \centering
    \includegraphics[width=\linewidth]{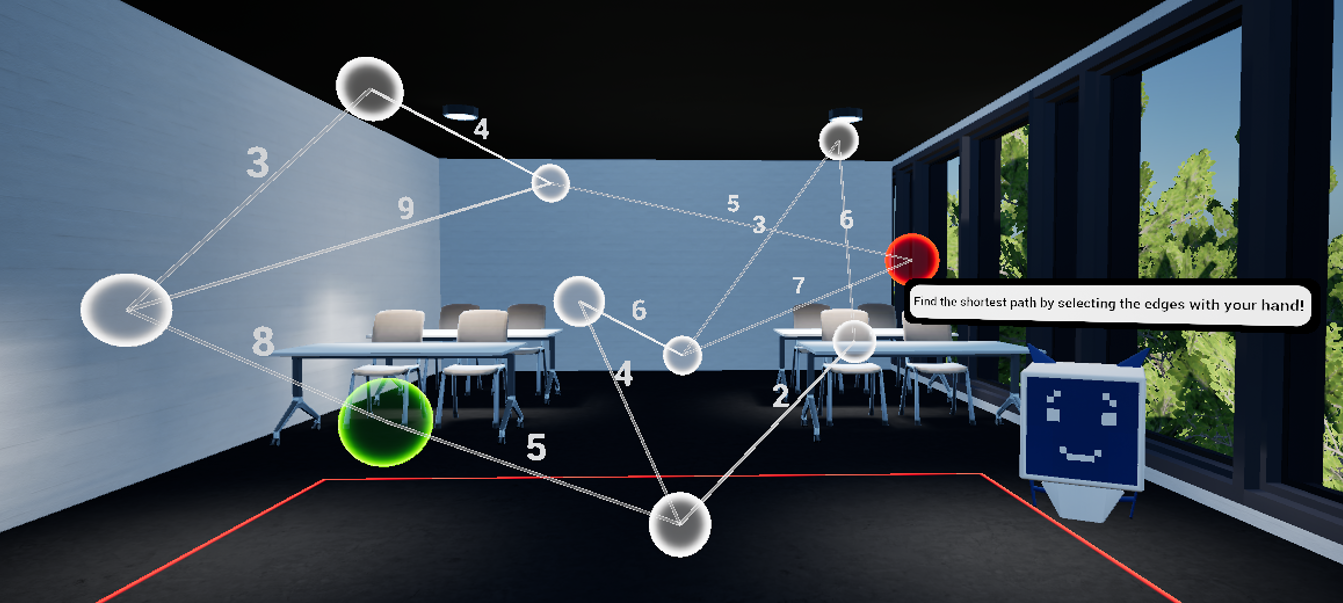}
    \caption{The Shortest Path mini-game: a weighted graph displayed in 3D space.}
    \label{fig:shortest}
\end{figure}

This mini-game introduces the players to the lectures Discrete Structures and Fundamentals of Algorithms. As illustrated in Figure \ref{fig:shortest}, the player is presented with a 3D network of 10 nodes connected by weighted edges. One node is colored green (start) and another red (goal). The player must interact with the edges to trace a path with the lowest possible total weight.

\textit{Learning outcome:} This mini-game introduces nodes, edges, edge weights, as well as the optimization problem of finding a minimum-cost route. These ideas underpin algorithms such as Dijkstra's and are central to several courses in the program.

\subsection{SQL Query -- Data Breach}

A more advanced mini-game presents a virtual monitoring console that displays a relational table \texttt{SensorReadings}, with columns for reading identifiers, station names, sensor types, measured values, and status codes. 

\begin{figure}[h!]
    \centering
    \includegraphics[width=\linewidth]{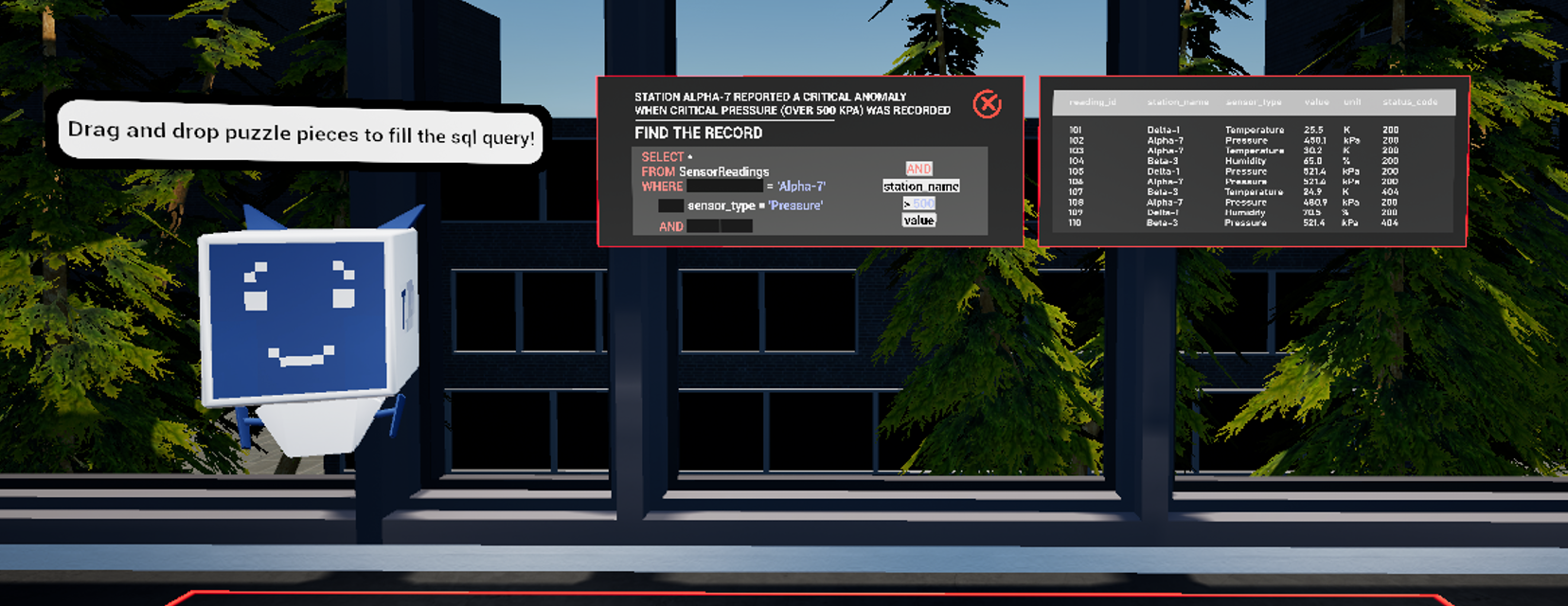}
    \caption{The SQL Query mini-game: constructing a multi-condition \texttt{WHERE} clause on a virtual console.}
    \label{fig:sql}
\end{figure}

Figure \ref{fig:sql} illustrates how a narrative prompt by TUMi states that station Alpha-7 reported a critical pressure anomaly exceeding 500\,kPa, and the player must construct the SQL query that isolates the offending record by filling in the blank slots with the correct column names, comparison operators, and logical connectives. 

The intended solution is:

\begin{verbatim}
SELECT * FROM SensorReadings
WHERE station_name = 'Alpha-7'
  AND sensor_type  = 'Pressure'
  AND value > 500;
\end{verbatim}

Submitting the correct query returns the single matching row (\texttt{reading\_id 106}, \texttt{value 521.4\,kPa}, \texttt{status 200}), giving the player immediate confirmation.

\textit{Learning outcome:} Players practice composing multi-condition \texttt{WHERE} clauses, reinforcing how relational filtering works. The sci-fi framing keeps the exercise engaging while teaching a skill directly assessed in the databases course.

Table~\ref{tab:alignment} summarizes how each mini-game maps to one or more courses in the Information Engineering program and the primary competencies it targets with a defined difficulty level based on the content of the mini-game and the associated lecture. 

\begin{strip}
\centering
\captionof{table}{Synthesis of TUMSphere mini-games: curriculum mapping, interaction mechanics, and targeted competencies.}
\small
\begin{tabular}{@{}p{2.3cm}p{3cm}p{4.7cm}p{4.7cm}p{1.7cm}@{}}
\toprule
\textbf{Mini-Game} & \textbf{Linked Course(s)} & \textbf{VR Interaction Mechanic}  & \textbf{Key Competencies} & \textbf{Difficulty} \\
\midrule
First Contact & Intro to Programming & Grab-and-snap word blocks into ordered slots & Syntax awareness, statement ordering & Introductory \\
Fix The Elevator & Computer Architecture & Drag cables to color-matched sockets  & Signal matching, hardware-level reasoning & Introductory \\
Fix The Code & Intro to Programming, Computer Architecture & Select tokens from a set and place them into code blanks  & Variable handling, arithmetic logic, debugging & Introductory \\
Knight's Tour & Discrete Structures, Algorithms & Point-and-select squares on a configurable board  & Graph traversal, Hamiltonian paths, backtracking & Intermediate \\
Shortest Path & Discrete Structures, Algorithms & Interact with edges of a weighted 3D graph  & Weighted graphs, path optimization & Intermediate \\
SQL Query & Intro to Databases & Fill blank slots in a query template on a virtual console  & Relational filtering, multi-condition queries & Advanced \\
\bottomrule
\end{tabular}
\label{tab:alignment}
\end{strip}

\section{Design Considerations}\label{sec:design}

Several cross-cutting design decisions, refined throughout the development process and confirmed during the final analysis, shaped all of the mini-games:

\paragraph{Blueprint-based development.} The game was implemented using Unreal Engine~5's visual scripting system (Blueprints), which allowed rapid iteration and made it straightforward to share reusable logic components across all mini-games.

\paragraph{Mixed-dimensionality assets.} Puzzles that revolve around textual or symbolic reasoning (e.g., code completion, SQL construction) use flat 2D interface panels optimized for readability. In contrast, spatially oriented tasks (e.g., elevator wiring, the knight's board) employ full 3D models to exploit the depth cues VR naturally provides.

\paragraph{Graduated difficulty.} The ordering of mini-games deliberately follows the semester sequence of the degree program. Players encounter simpler, syntax-level tasks first and progress toward challenges that require more logical reasoning.

\paragraph{Immediate feedback.} Each mini-game provides instant visual or auditory confirmation of correct and incorrect actions, color changes, glow effects, and score increments, so that players can self-correct without external guidance.

\paragraph{Comfort and cybersickness mitigation.} All locomotion relies on teleportation rather than continuous movement, reducing the vestibular mismatch that is a primary cause of VR-induced discomfort. Mini-games are designed to take place within a confined spatial area so players can remain mostly stationary while interacting, and environmental transitions between challenges use brief fade-to-black screens rather than rapid camera motions.

\paragraph{Consistent interaction language.} A uniform set of input gestures is reused across all mini-games: grip to grab, trigger to confirm, and point to select. This consistency means that the motor skills taught in the tutorial transfer directly to subsequent challenges, minimizing the cognitive overhead of learning new controls.

\paragraph{Modular architecture.} Each mini-game is encapsulated as an independent Blueprint actor with a standardized interface for initialization, scoring, and completion callbacks. This modular design decouples game logic from the campus environment, allowing the addition, removal, or reordering of challenges without modifying unrelated parts of the codebase. This property directly supports the extensibility of curriculum mapping.

\paragraph{Narrative coherence.} Rather than presenting the mini-games as disconnected exercises, TUMSphere embeds them within a light narrative arc guided by the TUMi companion. TUMi contextualizes each challenge with a brief spoken introduction that relates the task to its real-world academic counterpart and offers encouragement upon completion. This narrative thread helps maintain player motivation across the full sequence and reinforces the connection between gameplay and coursework.

\paragraph{Accessibility considerations.} Interface panels use high-contrast color schemes and sufficiently large font sizes to remain legible at typical VR reading distances. Interactive elements are placed within the comfortable reach envelope defined by Meta's design guidelines, and all color-coded feedback is supplemented with shape or animation cues so that color-vision-deficient players can still interpret the game state correctly. Following established VR design guidelines \cite{mehmedova2025virtual}, TUMSphere requires controller-based input rather than hand tracking throughout the experience.

Table~\ref{tab:design} summarizes each design consideration, its scope, and the underlying rationale, serving as a practical reference for ongoing development and for researchers seeking to adopt or extend this approach.

\begin{strip}
\centering
\captionof{table}{Summary of cross-cutting design considerations in TUMSphere.}
\small
\begin{tabular}{@{}p{3cm}p{1.4cm}p{6.1cm}p{6.4cm}@{}}
\toprule
\textbf{Design Consideration} & \textbf{Scope} & \textbf{Implementation} & \textbf{Rationale} \\
\midrule
Blueprint-based development & Development & All game logic authored in UE5 Blueprints with shared reusable components. & Enables rapid iteration without C++ expertise; lowers the barrier for an academic team. \\
Mixed-dimensionality assets & Content & 2D panels for text/symbolic puzzles; full 3D models for spatial tasks. & Matches asset format to task type, optimizing readability and depth perception respectively. \\
Graduated difficulty & Pedagogy & Mini-games are ordered to mirror the semester sequence of the degree program. & Ensures a smooth learning ramp-up aligned with real academic expectations. \\
Immediate feedback & UX & Color changes, glow effects, and score increments on every action. & Supports self-correction without external guidance; reinforces correct reasoning. \\
Comfort \& cybersickness mitigation & UX & Teleportation locomotion, confined play areas, fade-to-black transitions. & Reduces vestibular mismatch and motion-induced discomfort. \\
Consistent interaction language & UX & Uniform grip/trigger/point gesture set reused across all mini-games. & Minimizes control-learning overhead; lets players focus on academic content. \\
Modular architecture & Engineering & Each mini-game is an independent Blueprint actor with a standardized interface. & Decouples game logic from the environment, simplifying the addition or reordering of challenges. \\
Narrative coherence & Pedagogy & TUMi introduces each challenge with context, linking gameplay to coursework. & Maintains motivation and reinforces the connection between play and academic material. \\
Accessibility & UX & High-contrast UI, comfortable reach placement, shape/animation cues alongside color coding. & Ensures usability for players with varying visual abilities and physical reach. \\
\bottomrule
\end{tabular}
\label{tab:design}
\end{strip}

\section{Evaluation}\label{sec:eval}

A pilot study was conducted to assess TUMSphere's usability, engagement potential, and preliminary educational effectiveness.

\subsection{Study Design and Participants}

A within-subjects evaluation was conducted with $N = 18$ volunteers (11 male, 7 female; age range 18--24, $M = 19.8$, $SD = 2.4$). Participants were recruited from the TUM Heilbronn student body; none had previously used the application. To control for prior expertise, participants self-reported their VR experience on a 5-point scale (1 = \textit{no experience}, 5 = \textit{frequent user}); the sample mean was $2.3$ ($SD = 1.1$), indicating a predominantly novice-to-intermediate cohort.

Each session lasted approximately 40 minutes and followed a fixed protocol: (i)~a brief demographic questionnaire and a 10-item \textit{pre-test} covering the academic topics addressed by the mini-games; (ii)~a full play-through of TUMSphere, including the tutorial and all six mini-games; (iii)~the same 10-item \textit{post-test} administered immediately after gameplay; and (iv)~a post-session questionnaire comprising the System Usability Scale (SUS)~\cite{lewis2018system}, a 7-item Likert-scale engagement subscale adapted from the User Engagement Scale (UES)~\cite{o2018practical}, and the Simulator Sickness Questionnaire (SSQ)~\cite{kennedy1993simulator}. All participants provided informed consent, as the experiment was not invasive and we did not collect or store personal data, so we did not require Ethical Committee approval.

\subsection{Instruments and Metrics}

\paragraph{Knowledge gain.} The pre- and post-tests each contained 10 multiple-choice items (two per academic topic: programming syntax, hardware debugging, code completion, graph theory, and relational databases), scored on a 0--10 scale. Knowledge gain was computed as $\Delta = \text{post} - \text{pre}$.

\paragraph{Usability.} The standard 10-item SUS questionnaire yields a composite score between 0 and 100. A score above 68 is generally considered above average~\cite{lewis2018system}.

\paragraph{Engagement.} Seven items from the UES short form (focused attention, perceived usability, aesthetic appeal, and reward) were rated on a 5-point Likert scale (1 = \textit{strongly disagree}, 5 = \textit{strongly agree}). A composite engagement score was computed as the mean across all seven items.

\paragraph{Simulator sickness.} The SSQ produces a Total Severity score from 16 symptom items rated on a 4-point scale (0 = \textit{none}, 3 = \textit{severe}). Total Severity scores below 10 are generally classified as negligible~\cite{kennedy1993simulator}.

\paragraph{Task performance.} For each mini-game, the system logged the completion time (in minutes and seconds), the number of attempts before success, and a binary completion flag.

\subsection{Statistical Methods}

Given the small sample size ($N = 18$), non-parametric tests were preferred throughout the analysis. The Wilcoxon signed-rank test was used to compare pre- and post-test scores and to assess within-subject differences across mini-game completion times. Effect sizes were reported as $r = Z / \sqrt{N}$, where $Z$ is the Wilcoxon test statistic and $N$ the number of observations~\cite{fritz2012effect}. Spearman's rank-order correlation ($\rho$) was used to examine relationships between self-reported VR experience and task performance. Internal consistency of the engagement subscale was assessed using Cronbach's $\alpha$. All tests were conducted at $\alpha = .05$ (two-tailed).

\section{Results}\label{sec:results}

\subsection{Knowledge Gain}

Pre-test scores ranged from 2 to 7 ($Mdn = 4.0$, $M = 4.22$, $SD = 1.40$), while post-test scores ranged from 5 to 10 ($Mdn = 7.0$, $M = 6.89$, $SD = 1.28$). A Wilcoxon signed-rank test indicated that the improvement was statistically significant ($Z = -3.64$, $p < 0.001$, $r = 0.86$), representing a large effect. All 18 participants improved or maintained their score; no participant scored lower on the post-test. Table~\ref{tab:knowledge} provides a per-topic breakdown.

\begin{table}[h]
\centering
\caption{Pre- and post-test scores by academic topic (each scored 0--2). Values are $Mdn$ ($M \pm SD$).}
\small
\begin{tabular}{@{}p{3cm}cc@{}}
\toprule
\textbf{Topic} & \textbf{Pre-test} & \textbf{Post-test} \\
\midrule
Programming syntax    & 1.0 (0.89 $\pm$ 0.58) & 2.0 (1.56 $\pm$ 0.51) \\
Hardware debugging    & 1.0 (0.94 $\pm$ 0.64) & 1.5 (1.44 $\pm$ 0.51) \\
Code completion       & 1.0 (0.83 $\pm$ 0.62) & 1.0 (1.33 $\pm$ 0.59) \\
Graph theory          & 0.5 (0.72 $\pm$ 0.67) & 1.0 (1.22 $\pm$ 0.55) \\
Relational databases  & 1.0 (0.83 $\pm$ 0.71) & 2.0 (1.33 $\pm$ 0.59) \\
\bottomrule
\end{tabular}
\label{tab:knowledge}
\end{table}

The largest absolute gains were observed in programming syntax and relational databases. At the same time, graph theory showed the smallest improvement---consistent with the higher difficulty reported for the Knight's Tour and Shortest Path mini-games.

\subsection{Usability}

The SUS composite score across all participants was $M = 76.4$ ($SD = 9.2$, $Mdn = 78.75$), placing TUMSphere in the ``good'' usability band (above the 68-point benchmark). Individual scores ranged from 57.5 to 92.5. No participant scored below 50, and 14 of 18 (77.8\%) scored above 70.

\subsection{Engagement}

The seven-item engagement subscale demonstrated good internal consistency ($\alpha = 0.84$). The composite engagement score was $M = 4.21$ ($SD = 0.53$) on the 5-point scale, with individual item means ranging from 3.89 (``I lost track of time while playing'') to 4.56 (``The VR environment was visually appealing''). Table~\ref{tab:engagement} reports the item-level descriptive statistics.

\begin{table}[h]
\centering
\caption{Engagement subscale item-level results ($N = 18$). Scale: 1--5.}
\small
\begin{tabular}{@{}p{6.5cm}cc@{}}
\toprule
\textbf{Item} & $M$ & $SD$ \\
\midrule
I felt focused during the mini-games.            & 4.28 & 0.57 \\
I lost track of time while playing.              & 3.89 & 0.83 \\
The VR environment was visually appealing.        & 4.56 & 0.51 \\
The challenges felt rewarding to complete.        & 4.33 & 0.59 \\
I was motivated to complete all mini-games.       & 4.22 & 0.65 \\
The controls were easy to use.                    & 4.11 & 0.68 \\
I would recommend this experience to others.      & 4.06 & 0.73 \\
\bottomrule
\end{tabular}
\label{tab:engagement}
\end{table}

\subsection{Simulator Sickness}

SSQ Total Severity scores were low across the sample ($M = 7.1$, $SD = 5.3$, $Mdn = 5.6$), falling below the 10-point threshold commonly used to classify negligible sickness~\cite{kennedy1993simulator}. Only 2 of 18 participants (11.1\%) reported scores above 15, and no participant reported severe symptoms on any individual item. These results suggest that comfort-oriented design decisions (teleportation locomotion, confined play areas, and fade-to-black transitions) minimize VR-induced discomfort.

\subsection{Task Performance}

Table~\ref{tab:performance} summarizes the completion time and attempt count for each mini-game. All 18 participants completed the full mini-game sequence. The introductory-level challenges (First Contact, Fix The Elevator, Fix The Code) exhibited short completion times and low attempt counts. The intermediate-level Knight's Tour showed the highest variability in both time ($SD = 68.2$\,s) and attempts ($SD = 1.9$), confirming qualitative feedback that this puzzle had the steepest difficulty curve. A Friedman test across the six mini-games revealed a significant effect of mini-game on completion time ($\chi^2(5) = 72.3$, $p < 0.001$), consistent with the intended graduated difficulty design.

\begin{table}[h]
\centering
\caption{Task performance per mini-game ($N = 18$). Time in seconds; attempts counted as discrete retries.}
\small
\begin{tabular}{@{}p{2.3cm}cc@{}}
\toprule
\textbf{Mini-Game} & \textbf{Time ($M \pm SD$)} & \textbf{Attempts ($M \pm SD$)} \\
\midrule
First Contact    & 42.3 $\pm$ 15.1 & 1.2 $\pm$ 0.4 \\
Fix The Elevator & 35.7 $\pm$ 12.8 & 1.1 $\pm$ 0.3 \\
Fix The Code     & 78.4 $\pm$ 28.6 & 1.6 $\pm$ 0.8 \\
Knight's Tour    & 195.2 $\pm$ 68.2 & 3.4 $\pm$ 1.9 \\
Shortest Path    & 124.6 $\pm$ 41.3 & 2.1 $\pm$ 1.0 \\
SQL Query        & 156.8 $\pm$ 52.7 & 2.5 $\pm$ 1.2 \\
\bottomrule
\end{tabular}
\label{tab:performance}
\end{table}

\subsection{Correlation Between VR Experience and Performance}

Spearman's correlation revealed a moderate negative relationship between self-reported VR experience and total completion time ($\rho = -0.52$, $p = 0.027$), indicating that participants with more VR familiarity tended to finish the full sequence faster. However, VR experience was not significantly correlated with knowledge gain ($\rho = 0.18$, $p = 0.47$), suggesting that prior VR exposure did not confer an advantage in learning the academic content itself.

\section{Conclusion}\label{sec:conclusion}

The pilot evaluation yielded converging evidence that TUMSphere's curriculum-driven mini-games are both usable and educationally effective as a first encounter with the Information Engineering program. Participants showed a statistically significant increase in knowledge from pre- to post-test, with the largest improvements in programming syntax and relational databases. System usability was rated as ``good'' (SUS $M = 76.4$), engagement was consistently high across all seven subscale items ($M = 4.21 / 5$), and simulator sickness remained negligible (SSQ $M = 7.1$). Task performance data confirmed the intended graduated difficulty: introductory mini-games were completed quickly with few retries, while the Knight's Tour showed the greatest variability in time and attempts, suggesting a need for additional in-game scaffolding at the intermediate level.

The mini-games in TUMSphere illustrate a practical method for embedding university-level subject matter directly into interactive VR mechanics. By aligning each challenge with a specific course and ordering them in increasing order of academic complexity, the system offers students a playful yet substantive preview of their studies. The approach is inherently extensible: additional courses can be represented by designing new mini-games that follow the same curriculum-mapping principle, making TUMSphere a scalable template for gamified academic outreach.

\section{Future Work}\label{sec:future}

Several directions are planned to extend both the breadth and the rigor of TUMSphere.

\paragraph{Curriculum expansion.} Upcoming development cycles will introduce challenges for courses not yet represented, such as signal processing, machine learning, and software engineering. Each new mini-game will follow the same design methodology, so that the full-semester arc from the first to the final year is eventually playable.

\paragraph{Adaptive difficulty.} The pilot study revealed notable variation in how quickly participants completed the more demanding puzzles. A natural next step is to implement an adaptive difficulty system that adjusts parameters in real time — for instance, reducing the board size in the Knight's Tour or providing incremental hints in the SQL challenge — based on the player's observed performance and response latency. Such a mechanism would help prevent both frustration for novice users and under-stimulation for more experienced ones.

\paragraph{Multiplayer and collaborative modes.} All mini-games are currently single-player experiences. Introducing cooperative or competitive multiplayer modes — for example, two players collaboratively constructing an SQL query or racing to find the shortest path on the same graph — would open the door to peer learning dynamics and could strengthen the social dimension of the VR campus.

\paragraph{Larger-scale empirical evaluation.} A follow-up controlled study with a larger and more diverse cohort is planned, incorporating a between-subjects comparison with a traditional orientation format, physiological measures of cognitive load (e.g., pupillometry), and delayed retention tests to assess longer-term learning effects.

\paragraph{Accessibility and platform portability.} To broaden access, future iterations will explore deployment on additional headsets as well as a reduced-fidelity desktop or web-based fallback for users without VR hardware.

\section*{Acknowledgments}
The authors used AI-based tools solely for grammar checking and voice correction during the preparation of this manuscript. No AI was used for content generation, data analysis, or result interpretation.  

\bibliographystyle{ieeetr}
\bibliography{00References}

\end{document}